\documentclass[journal,10pt]{IEEEtran}
\usepackage{epsfig}
\usepackage{diagbox}
\usepackage{amsmath}
\usepackage{amsfonts}
\usepackage{array}
\usepackage{amssymb}

\usepackage{clrscode3e}
\usepackage[compress]{cite}
\usepackage{algorithm}
\usepackage{color}
\usepackage{algorithmicx}
\usepackage{algpseudocode}
\usepackage{graphicx}
\usepackage{subfig}
\usepackage{url}
\usepackage{changepage}
\usepackage{booktabs}
\usepackage{multirow}
\usepackage{float}
\usepackage{array}
\usepackage{multirow}
\usepackage{mdwmath}
\usepackage{mathrsfs}
\usepackage{mdwtab}
\usepackage{graphics}
\usepackage{textcomp}
\usepackage{rotating}
\usepackage{epstopdf}

\begin{document}
\title{De-END: Decoder-driven Watermarking Network}
\providecommand{\keywords}[1]{\textbf{\textit{Index terms---}}#1}

\author{Han~Fang, Zhaoyang~Jia, Yupeng~Qiu, Jiyi~Zhang, Weiming~Zhang and Ee-Chien~Chang
\thanks{Han Fang, Yupeng~Qiu, Jiyi~Zhang and Ee-Chien~Chang is with School of Computing, National University of Singapore, Singapore. Zhaoyang~Jia and Weiming Zhang are all with CAS Key Laboratory of Electromagnetic Space Information, University of Science and Technology of China, Hefei, 230026, China. (e-mail: fanghan@nus.edu.sg, zhangwm@ustc.edu.cn)}
% \thanks{This work was supported in part by the Natural Science Foundation of China
% under Grant 62072421, 62002334, 62121002 and U20B2047, Anhui Science Foundation of China under Grant 2008085QF296, and by Exploration Fund Project of University of Science and Technology of China under Grant YD3480002001.}
}
\maketitle

\begin{abstract}
With recent advances in machine learning, researchers are now able to solve traditional problems with new solutions. In the area of digital watermarking, deep-learning-based watermarking technique is being extensively studied. Most existing approaches adopt a similar encoder-driven scheme  which we name \textbf{END} (Encoder-NoiseLayer-Decoder) architecture. In this paper, we revamp the architecture and creatively design a decoder-driven watermarking network dubbed \textbf{De-END} which greatly outperforms the existing \textbf{END}-based methods. 
The motivation for designing \textbf{De-END} originated from the potential drawback we discovered in \textbf{END} architecture: The encoder may embed redundant features that are not necessary for decoding, limiting the performance of the whole network. We conducted a detailed analysis and found that such limitations are caused by unsatisfactory coupling between the encoder and decoder in \textbf{END}. 
\textbf{De-END} addresses such drawbacks by adopting a \textbf{De}coder-\textbf{E}ncoder-\textbf{N}oiselayer-\textbf{D}ecoder architecture. In \textbf{De-END}, the host image is firstly processed by the decoder to generate a latent feature map instead of being directly fed into the encoder. This latent feature map is concatenated to the original watermark message and then processed by the encoder. This change in design is crucial as it makes the feature of encoder and decoder directly shared thus the encoder and decoder are better coupled.
We conducted extensive experiments and the results show that this framework outperforms the existing state-of-the-art (SOTA) \textbf{END}-based deep learning watermarking both in visual quality and robustness. On the premise of the same decoder structure, the visual quality (measured by PSNR) of \textbf{De-END} improves by 1.6dB (45.16dB $\rightarrow$ 46.84dB), and extraction accuracy after JPEG compression (QF=50) distortion outperforms more than 4\% ($94.9\%\rightarrow 99.1\%$).
\end{abstract}


\IEEEpeerreviewmaketitle

\section{Introduction}\label{Introduction}

Deep neural network (DNN) has powerful feature extraction capability. Such capability makes it a good replacement of traditional algorithms in many applications, with better performance or efficiency. 

Watermarking technique is an important mechanism for copyright protection and leak source tracing \cite{tsui2008color,nezhadarya2011robust,urvoy2014perceptual}. A watermarking scheme has to guarantee two properties: robustness and transparency. To satisfy these properties, image features have to be first extracted and then embedded into corresponding coefficients. Traditionally, the feature extraction is achieved by handcrafted operations (\textit{e.g.} discrete Fourier transformation \cite{tsui2008color}, discrete cosine transformation \cite{ko2020robust} and discrete wavelet transformation \cite{daren2001dwt}).

\begin{figure}[]
\begin{center}
    \subfloat[\textbf{END} architecture.]{
    \label{TeaserEND}
        {\centering\includegraphics[width=0.8\linewidth]{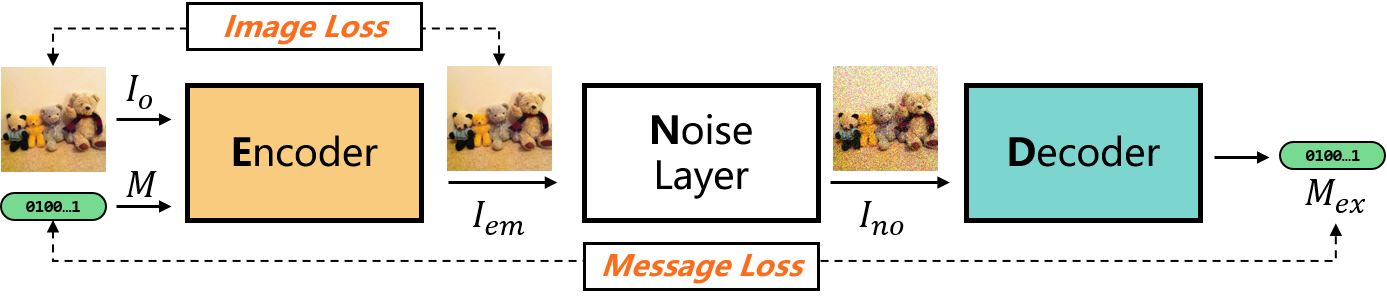}
        }
    }\\
    \subfloat[\textbf{De-END} architecture.]{
    \label{TeaserDEEND}
        {\centering\includegraphics[width=1\linewidth]{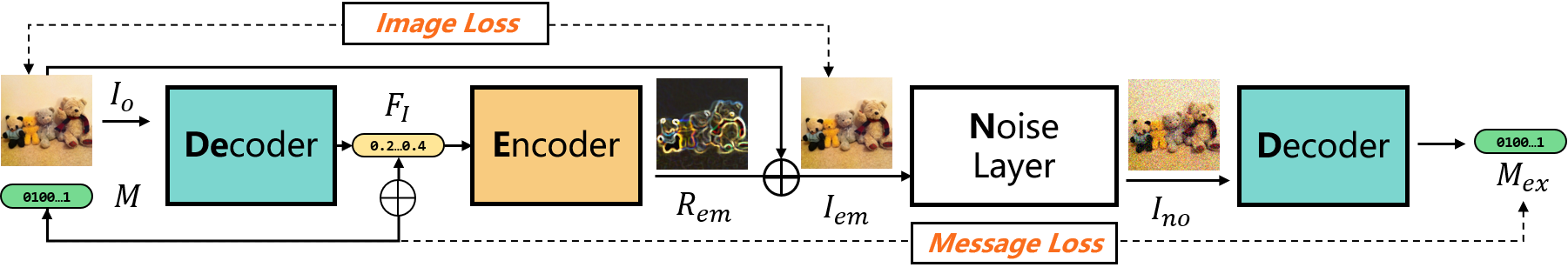}
        }
    }
\end{center}
\caption{The difference between \textbf{END} architecture and \textbf{De-END} architecture.}
\label{Teaser}
\end{figure}

Recently, many deep-learning-based watermarking schemes \cite{mun2017robust,mun2019finding,ahmadi2020redmark,kandi2017exploring,tancik2019stegastamp,wengrowski2019light,zhang2020udh,zhang2020viscode} proposed to replace the feature extractions process with DNN which effectively serves as a better embedding and extracting mechanism when training with the proper loss function and architecture. 
\\ \\
\noindent
{\bf END architecture.}\ \ 
The existing DNN-based watermarking network adopts an auto-encoder-like architecture as the main backbone which contains an encoder, a noise layer and a decoder, as shown in Fig \ref{TeaserEND}. In this paper, we name such architecture as \textbf{END}.
In \textbf{END} architecture, encoder, noise layer and decoder are cascaded serially. The encoder tries to embed the watermark into the host image and generate the watermarked image. The noise layer distorts the watermarked image. And the decoder aims to extract the watermark from the distorted image.
% encoder tries to embed the watermark into the host image with as less visual distortion as possible; noise layer adds some noise into the watermarked image in order to guarantee the robustness; decoder aims to extract the watermark features and conduct the decoding process. 
\\ \\
\captionsetup{font={footnotesize}}
\noindent
{\bf Limitations of END.}\ \ 
Although \textbf{END} can be jointly trained, it is not the optimal architecture for watermarking system. In fact, we discovered a potential drawback in this architecture, that is: the encoder might embed some redundant features into the host image and such redundant features could limit the performance of the architecture.

% \textbf{END} follows the rule of encoder-driven that encoder embeds a feature and decoder extract the encoded feature adaptively, but encoder will not directly influenced by decoder feature. So this will lead to feature redundant that not all the encoded feature are necessary for decoding. And the redundant feature will greatly influence the performance.

We believe that such drawback comes from the imperfect coupling of encoder and decoder. Because
% the core reason for this drawback is that encoder and decoder are not well coupled, so the feature of encoder and decoder cannot be effectively shared. 
in \textbf{END} architecture, the input of the encoder is the host image and watermark signal, which is independent to the decoder. But the input of the decoder is directly determined by the encoder, 
% The only connection from decoder to encoder is message loss in parameter updating. 
which makes \textbf{END} follows the encoder-driven rule that encoder embeds a feature and decoder extracts the encoded feature adaptively, and the features of the decoder cannot be well shared with the encoder. So training in this way will lead to differences in feature expression.
\\ \\
\noindent
{\bf De-END architecture.}\ \
To address drawbacks of \textbf{END}, in this paper, we propose \textbf{De-END}, as shown in Fig. \ref{TeaserDEEND}, a decoder-driven watermarking network. Specifically, \textbf{De-END} adopts a \textbf{De}coder-\textbf{E}ncoder-\textbf{N}oiselayer-\textbf{D}ecoder architecture, where the two decoder share the same parameters. In \textbf{De-END}, the host image is first fed into the decoder and outputs a latent feature that represents the decoder's analysis of the host image. Then the latent feature is concatenated with the watermarked signal, and the concatenated feature will be sent to the encoder. The following noise layer and decoder are same as \textbf{END} settings. Based on this structure, the output of the decoder could determine the input of the encoder, and the input of the decoder is also determined by the output of the encoder, which ensures that the feature of encoder and decoder can be effectively shared, and encoder could be better coupled with decoder.
\\ \\
\noindent
{\bf Contributions.}\ \
The main contributions of the proposed scheme are summarized as follows:
\begin{enumerate}%[leftmargin=*,nosep]
    \item We discover the potential drawback of the existing \textbf{END} architecture and analyze the key reason for this drawback. We hope the proposed analysis will benefit the follow-up watermarking scheme.
    \item We propose \textbf{De-END}, a novel decoder-driven watermarking network architecture which could effectively couple the encoder and decoder. This architecture leads a potential way in designing high-performance watermark networks.
    \item Various experiments indicate the superior performance of the proposed \textbf{De-END} architecture compared with the state-of-the-art DNN-based watermarking schemes both on visual quality and robustness. 
\end{enumerate}

\section{Related Work}\label{RelatedWork}
\subsection{Traditional watermarking scheme}
Traditional watermarking scheme \cite{fang2020deep,mellimi2021fast} follows the encoder-driven rules. 
% The design complexity of the algorithm is concentrated more on the embedding side, and the extraction procedure is only the inverse process of the embedding procedure. 
In order to meet the robustness requirement, the watermark is often embedded into the coefficients of transformed domain. And the choice of transformed domain depends on the qualitative analysis of the distortion. For example, to ensure the robustness of JPEG compression, traditional watermarking often embedded the watermark into the coefficients of discrete cosine transformation (DCT) domain \cite{fang2018screen}, since JPEG compression is carried out on DCT coefficients. The most commonly used transformation includes discrete cosine transformation \cite{langelaar2001optimal,fang2018screen}, discrete wavelet transformation (DWT) \cite{hu2019frame,gao2019dynamic} and discrete Fourier transformation (DFT) \cite{kang2011geometric,kang2010efficient}. Based on the different characteristics of these transformations, the target robustness of watermarking schemes can be effectively achieved. But since traditional methods only apply handcrafted features to perform the embedding and extraction process, they cannot well balance the visual quality and the robustness.
% \subsection{Watermarking scheme with DNN-based Decoder}
% In order to utilize the powerful feature extraction ability of DNN in watermarking scheme, some recent papers have proposed to use DNN-based decoder to replace the traditional handcrafted extraction algorithm. Fang \textit{et. al.} \cite{fang2020deep} proposed a DNN-based decoder combined with the traditional template-based embedding algorithm. By training with the distorted image blocks, the decoder can be trained to adapt to specific distortions. Mellimi \textit{et. al.} \cite{fang2020deep,mellimi2021fast} proposed a DNN-based extraction network to achieve better robustness. But this kind of method still uses the traditional DWT-based handcrafted embedding algorithm, which still belongs to encoder-driven watermark algorithm. Although applying DNN-based decoder can improve the robustness, the embedding algorithm is not well coupled with the decoder.

\subsection{DNN-based watermarking scheme}
Recently, the framework with DNN-based encoder and DNN-based decoder was proposed \cite{zhu2018hidden,liu2019a,jia2021mbrs,tancik2019stegastamp,wengrowski2019light,zhang2020udh,zhang2020viscode}. Zhu \textit{et. al.} \cite{zhu2018hidden} first proposed this framework and realize the common image processing robustness such as JPEG compression, Gaussian noise and so on. Tancik \textit{et. al.} \cite{tancik2019stegastamp} proposed a print-shooting noise layer to simulate the print-shooting process and further set it as the noise layer to train the whole network. As a result, the print-shooting robustness is guaranteed. Wengrowski \textit{et. al.} \cite{wengrowski2019light} proposed to use a CDTF network to simulate the screen-shooting process and regarded the well-trained CDTF network as the noise layer to guarantee the screen-shooting robustness. Liu \textit{et. al.} \cite{liu2019a} proposed a two-stage method to improve the robustness, they believed the decoder can be further fine-tuned to acquire stronger robustness against non-differentiable distortions. So they fix the pre-trained encoder and further trained the decoder only to improve the robustness. Jia \textit{et. al.} \cite{jia2021mbrs} proposed a mini-batch-based JPEG noise layer to improve the JPEG robustness, by alternatively training the network with ``real JPEG'' and ``simulated JPEG'' noise, the JPEG robustness can be greatly guaranteed. However, these methods all follow the \textbf{END} architecture which although build a connection with the encoder and the decoder, they cannot well couple the encoder and the decoder. 

\section{Analysis of \textbf{END} Architecture}
% \captionsetup{font={footnotesize}}
% \begin{figure}
%     \begin{minipage}[p]{1\linewidth}
%         \centering{
%             \includegraphics[width=0.9\linewidth]{samples/figures/ENDLOSS.png}
%         }
%     \end{minipage}%
% \caption{The framework of \textbf{END}. It consist of an encoder, a noise layer and a decoder. And the whole network is trained with image loss and message loss.}
% \label{END}
% \end{figure}
\captionsetup{font={footnotesize}}
\begin{figure}
    \begin{minipage}[p]{1\linewidth}
        \centering{
            \includegraphics[width=0.9\linewidth]{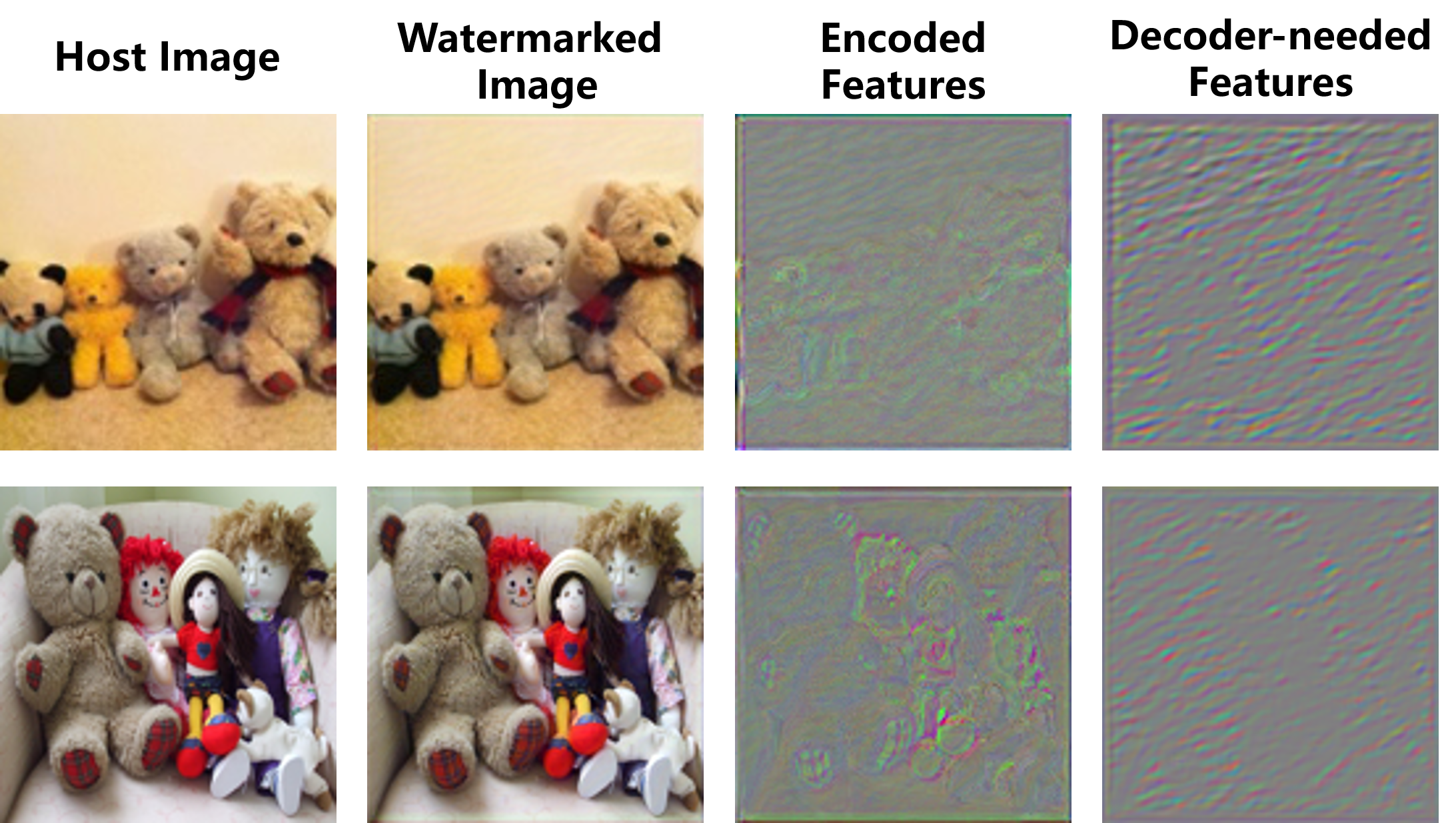}
        }
    \end{minipage}%
\caption{Two examples of images and corresponding features that are generated with HiDDen \cite{zhu2018hidden}.}
\label{ENDAnalysis}
\end{figure}
\captionsetup{font={footnotesize}}
\begin{figure*}
    \begin{minipage}[p]{1\linewidth}
        \centering{
            \includegraphics[width=1\linewidth]{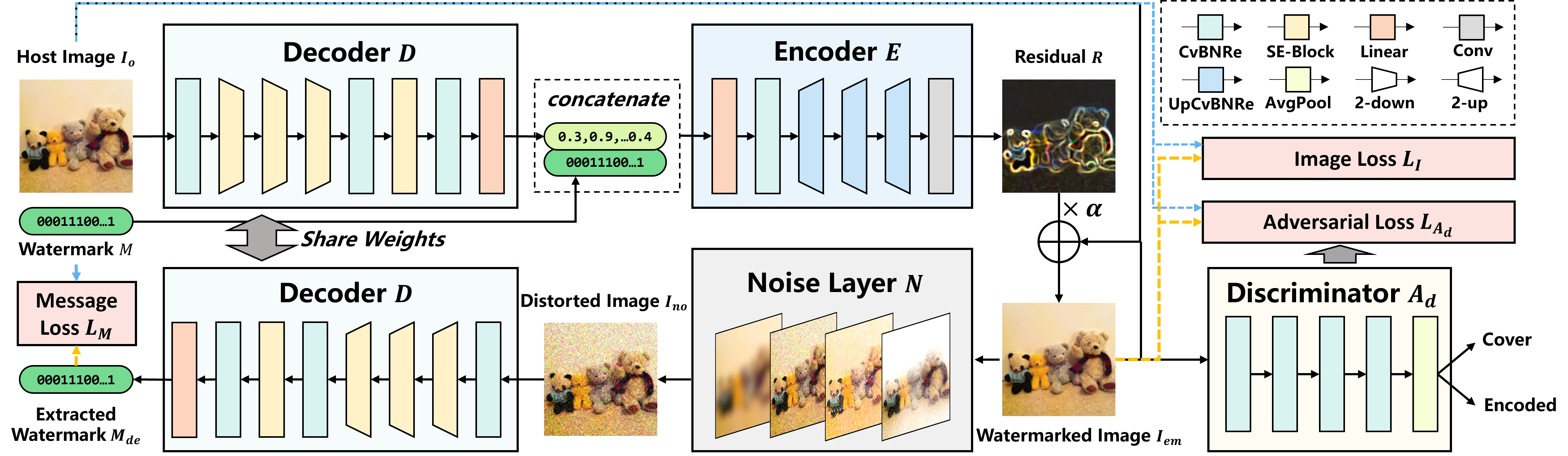}
        }
    \end{minipage}%
\caption{The framework of \textbf{De-END}. It consist of three main parts: the encoder, the noise layer and the decoder. The two decoders in the framework maintain the same parameters. And the whole framework can be trained end-to-end.}
\label{Framework}
\end{figure*}
Fig. \ref{TeaserEND} shows the typical framework with \textbf{END} architecture, which contains an encoder, a noise layer and a decoder. The whole network is optimized by the image loss and the message loss. In \textbf{END} architecture, the input of the encoder is host image and watermark message, and the output of encoder is the watermarked image. For decoder, the input is the distorted image and the output of decoder is the extracted message. Since the output of the encoder will directly determine the input of the decoder, the features of encoder can be directly shared to the decoder. But the features of decoder can only be shared indirectly through message loss to the encoder. So in \textbf{END}, feature sharing will be conducted in an unbalanced way.
% So the encoder and the decoder cannot be well coupled in \textbf{END} architecture.
% So such architecture is encoder-driven, which means the decoder tries to adapt to the encoder. 
% Therefore, for \textbf{END} architecture, encoder is not well coupled with decoder.
As a result, there will be a potential drawback, that is, encoder might embed redundant features which are useless for decoding into the host image. And the existence of redundant features will influence the visual quality of the watermarked image and the extraction accuracy of decoder.
% limit the performance of the watermarking network.

For better illustration, we visually show two examples of the encoded features and the decoder-needed features based on the method HiDDen \cite{zhu2018hidden} which utilizes \textbf{END} architecture. For the features that decoder needed, we generated it with the following operations: we feed the host image into the pre-trained decoder and calculate the MSE loss of the output and watermark messages. After that, we backpropagate the loss and take the gradient map on the host image as the features that decoder needed. The corresponding results are shown in Fig. \ref{ENDAnalysis}. $\widetilde{R}$ and $\widetilde{\nabla_{I}D}$ indicate the normalized encoded features and decoder-needed features respectively. It can be seen that the encoded features do not maintain high consistency with the decoder-needed features, there are redundant features being embedded, which indicates that the encoder is not well coupled with the decoder in \textbf{END} architecture.

In order to better couple the encoder and decoder, we should revamp the architecture to make sure that the features should be directly shared in both directions of the encoder and decoder. In this way, not only can the decoder adapt to the encoded features, but also the decoder-needed features can guide the encoder. Therefore, in this paper, we propose \textbf{De-END}, a decoder-driven watermarking network that can effectively couple the encoder and decoder.

\section{Proposed Architecture}
The framework of the proposed \textbf{De-END} is shown in Fig. \ref{Framework}, which consist of four main parts: (1) The decoder $D$ with parameters $\theta_D$; (2) the encoder $E$ with parameters $\theta_E$; (3) the noise layer $N$ and (4) the discriminator $Ad$ with parameters $\theta_{A_d}$. The working flow of whole framework can be described as: The host image $I_o\in \mathbb{R}^{C\times\ H\times W}$ is first fed into $D$. $D$ will output a latent feature $F\in \mathbb{R}^{L}$ which has the same size of watermark message $M\in\{0,1\}^{L}$ ($L$ indicates the length of $W$). Then $F$ and $M$ will be concatenated and further sent into $E$. The output of $E$ is a residual signal $R\in \mathbb{R}^{C\times\ H\times W}$ which will be weighted by a strength factor $\alpha$ and further superimposed on $I_o$ to generate the watermarked image $I_{em}\in \mathbb{R}^{C\times\ H\times W}$. After that, $I_{em}$ is distorted by the noise layer $N$ to generate the distorted image $I_{no}\in \mathbb{R}^{C\times\ H\times W}$. Finally, $D$ will try to decode the watermark $M_{de}\in\{0,1\}^{L}$ from $I_{no}$. Among them, $Ad$ is used for optimizing the visual quality of the watermarked image and will not participate in the encoding and decoding.
 
\subsection{Decoder}
There are two main purposes of the decoder in this framework. The first one is guiding the encoder, which aims to generate a latent image feature $F$ with the input of $I_o$. $F$ represents the expression of $I_o$ for $D$. After that, $F$ will be concatenated with $M$ and further sent to $E$. In this way, the input of $E$ is closely connected to the output of $D$, so that the features required by $D$ can be effectively and directly shared with $E$. The second purpose is being guided by the encoder, which tries to extract the encoded feature from $I_{no}$ and realize the accurate decoding. Since $I_{no}$ is determined by the features of $E$, achieving accurate decoding is equivalent to making $D$ adapt to the features of $E$. 
% In this way, $D$ is further coupled with $E$. 

The structure of the decoder $D$ is shown in Fig. \ref{Framework}. Specifically, one ``Conv-BN-ReLU'' block is first processed on $I_o$ to generate $64\times H\times W$ feature maps, then three ``SE-block'' \cite{hu2018squeeze} are applied to downsample the processed feature into a $64\times H/8\times W/8$ feature maps. After that, two ``Conv-BN-ReLU'' block, one ``SE-block'' and one ``Linear'' block are adopted to generate $F$ or $M_{de}$ with length $L$.

One object of $D$ is to minimize the difference between $M_{de}$ and the original watermark $M$ by updating $\theta_D$, which can be formulated by:
\begin{equation}
\mathcal{L}_{D}=M S E\left(M, M_{de}\right)=M S E\left(M, D\left(\theta_{D}, I_{n o}\right)\right)
\end{equation}

\subsection{Encoder}
Different from the traditional \textbf{END} architecture, the encoder in \textbf{De-END} only realizing the up-sampling process. It only takes the input of concatenated feature which is partially determined by $D$ and then outputs a residual $R$ which is further superimposed on $I_o$ to generate the watermarked image $I_{em}$. The structure of $E$ is shown in Fig. \ref{Framework}. Specifically, one ``Linear'' block is first adopted to resize the concatenated feature into a feature with length $H/8\times W/8$. Then the feature is reshaped and processed by a ``Conv-BN-ReLU'' block. After that, three ``Up-Conv-BN-ReLU''  blocks are taken to up-sample the feature into the size of $64\times H\times W$. Finally, one single ``Conv'' block is applied to generate the residual $R$ with size $C\times H\times W$. It is worth noting that a strength factor $\alpha$ is adopted in generating the final watermarked image $I_{em}$. And $\alpha$ is used for balancing the visual quality and the robustness.

The object of $E$ is to generate a better residual $R$ to minimize the difference between $I_{em}$ and the host image $I_o$ by updating $\theta_E$, which can be formulated by:
\begin{equation}
\mathcal{L}_{E}=M S E\left(I_o, I_{em}\right)=M S E\left(I_o, E\left(\theta_{E}, F,M\right)\times \alpha+I_o\right)
\end{equation}

\subsection{Noise Layer}
The noise layer $N$ is the key to realizing robustness. By setting different differentiable image processing operations in $N$, the watermarked image $I_{em}$ will be distorted into different versions. And the distorted image $I_{no}$ will be further decoded by the decoder. So the distortion adopted in training will determine the final robustness. Commonly used distortion include ``JPEG Compression'', ``Gaussian Noising'', ``Median Filtering'' and so on.

\subsection{Discriminator}
One essential purpose of the framework is generating a watermarked image with a given host image and a watermark. Such process is similar to Conditional-GAN \cite{mirza2014conditional}, which aims to generate an image with a given condition. So in order to generate the high-quality watermarked image, we adopt the structure of discriminator. The structure of discriminator $Ad$ is same as \cite{zhu2018hidden} proposed. Specifically, it consists of four ``Conv-BN-ReLU'' blocks and an ``AveragePooling'' block. The discriminator performs as an adversary of the generating process and tries to give a correct identification between $I_{em}$ and $I_o$, which is realized by updating $\theta_{Ad}$:
% \begin{equation}\label{eq_L_Ad}
% \mathcal{L}_{Dis}=log(1-Ad(\theta_{Ad},I_{em})) = log(1-Ad(\theta_{Ad},E\left( F,M\right)\times \alpha+I_o))
% \end{equation}
\begin{equation}\label{eq_L_Ad}
\begin{aligned}
\mathcal{L}_{D i s} &=log \left(1-A d\left(\theta_{A d}, I_{e m}\right)\right) \\
&=log \left(1-A d\left(\theta_{A d}, E(F, M) \times \alpha+I_{o}\right)\right)
\end{aligned}
\end{equation}
On the other hand, $\theta_E$ should be also updated to mislead the discriminator, which can be realized by minimizing:
\begin{equation}\label{eq_L_Ad}
\mathcal{L}_{Ad}=log(Ad(I_{em})) = log(Ad(E(\theta_{E},F,M)\times \alpha+I_o))
\end{equation}

\subsection{Loss Function}
The final loss of the whole network is consist of image loss, adversarial loss and decoding loss, which can be formulated by:
\begin{equation}
    \mathcal{L} = \lambda_1\mathcal{L}_E + \lambda_2\mathcal{L}_D + \lambda_3\mathcal{L}_{Ad}
\end{equation}
where $\lambda_1,\lambda_2,\lambda_3$ are set as $1,10,0.0001$ in the first 20 epoch and $10,1,0.0001$ in the rest epoch.
\captionsetup{font={footnotesize}}
\begin{figure*}
    \begin{minipage}[p]{1\linewidth}
        \centering{
            \includegraphics[width=0.9\linewidth]{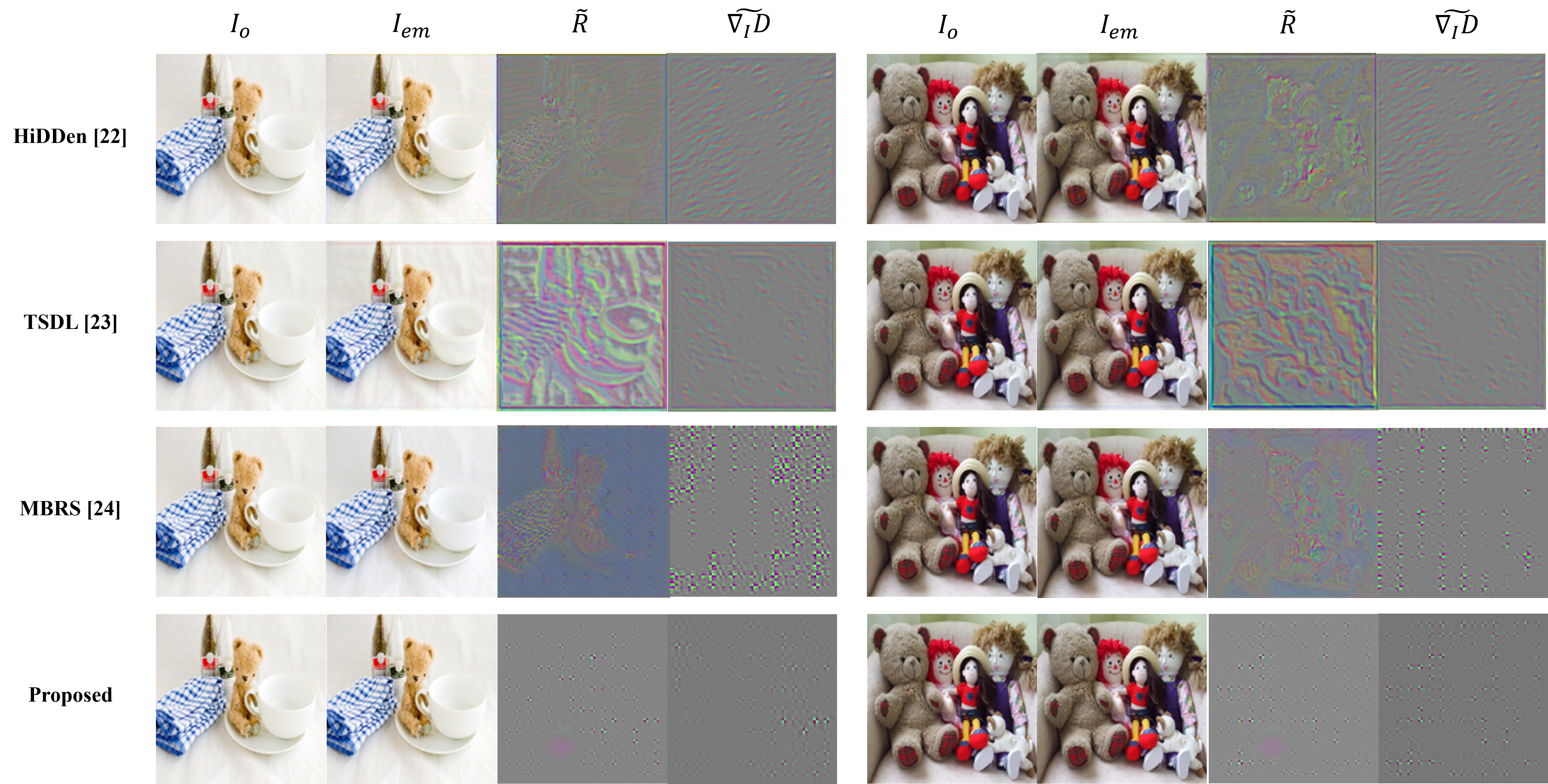}
        }
    \end{minipage}%
\caption{The images and corresponding features generated with different methods.}
\label{Coupling}
\end{figure*}
\section{Experimental Results}
In this section, we will first introduce the implementation details and the parameter selection of the proposed scheme. Then extensive experiments will be conducted to verify the robustness and visual quality of our method. Finally, more analysis of the proposed architecture will be provided to justify our design.
\subsection{Implementation Details}
The whole network is trained on COCO dataset \cite{lin2014microsoft}, and we randomly choose 10000 images as the training dataset. The framework is implemented by PyTorch \cite{collobert2011torch7:} and performed on NVIDIA RTX 2080ti. The size of the image $H$ and $W$ are both set as 128. The length $L$ is set as 64. Strength factor $\alpha$ is fixed as 1 in training. For neural network parameter optimization, we utilize Adam \cite{kingma2014adam:} with default hyperparameters. In testing experiments, we utilize the classical USC-SIPI image dataset \cite{SIPI} as our testing data.

For visual quality measurement, we use PSNR as the assessment metric. For robustness evaluation, we directly use the extraction accuracy of the watermark message as the assessment metric. We compare the proposed framework with three deep-learning-based framework HiDDen \cite{zhu2018hidden}, TSDL \cite{liu2019a} and MBRS \cite{jia2021mbrs}. The distortion we selected include ``Cropout'', ``Dropout'', ``Gaussian Noise'', ``Salt\&Pepper Noise'', ``Gaussian Blur'', ``Median Blur'' and ``JPEG Compression''. For each distortion, we train a specific network to better show the effectiveness of the architecture and the robustness of the framework. All the compared methods are all training with specified noise layers. Since the different noise layer will result in different encoder and decoder, the visual quality of the watermarked image and the extraction accuracy varies. So in this paper, we show the PSNR of the watermarked image and the extraction accuracy for each noise layer respectively.

\subsection{Coupling Measurement}
As aforementioned, \textbf{END} architecture has a potential drawback that the encoder and decoder may not be well coupled. And \textbf{De-END} can effectively handle this problem. So in this section, we will give two examples to visually compare the coupling of encoder and decoder with different schemes, as shown in Fig. \ref{Coupling}.

% \captionsetup{font={footnotesize}}
% \begin{figure*}[]
% \begin{center}
%     \subfloat[The image and features generated by HiDDen.]{
%     \label{TeaserEND}
%         {\centering\includegraphics[width=0.4\linewidth]{samples/figures/VQ_HiDDen.png}
%         }
%     }
%     \subfloat[The image and features generated by TSDL.]{
%     \label{TeaserEND}
%         {\centering\includegraphics[width=0.4\linewidth]{samples/figures/VQ_TSDL.png}
%         }
%     }
%     \\
%     \subfloat[The image and features generated by MBRS.]{
%     \label{TeaserEND}
%         {\centering\includegraphics[width=0.4\linewidth]{samples/figures/VQ_TSDL.png}
%         }
%     }
%     \subfloat[The image and features generated by proposed architecture.]{
%     \label{TeaserDEEND}
%         {\centering\includegraphics[width=0.4\linewidth]{samples/figures/VQ_Proposed.png}
%         }
%     }
% \end{center}
% \caption{The image and corresponding features generated with different methods.}
% \label{Coupling}
% \end{figure*}

The first column is the original host images, the second column represents the watermarked images, the third column indicates the normalized encoded features and the fourth column shows the normalized features that the decoder needed. The noise layer we choose is JPEG compression. As we can see in Fig. \ref{Coupling}, for the traditional \textbf{END} architecture, the embedded residual features are not highly consistent with the features required by the decoder. The encoder embeds more features than the decoder needed, which may reflect some image textures. This means that the encoder and decoder obtained by the \textbf{END} framework cannot be effectively coupled together, resulting in certain feature expression differences. However, for the proposed \textbf{De-END} architecture, the embedded features keep high consistency with the features decoder needed, which means that the encoder and decoder trained by this architecture can be better coupled.

\subsection{Visual Quality and Robustness}

\begin{figure*}
    \begin{minipage}[p]{1\linewidth}
        \centering{
            \includegraphics[width=1\linewidth]{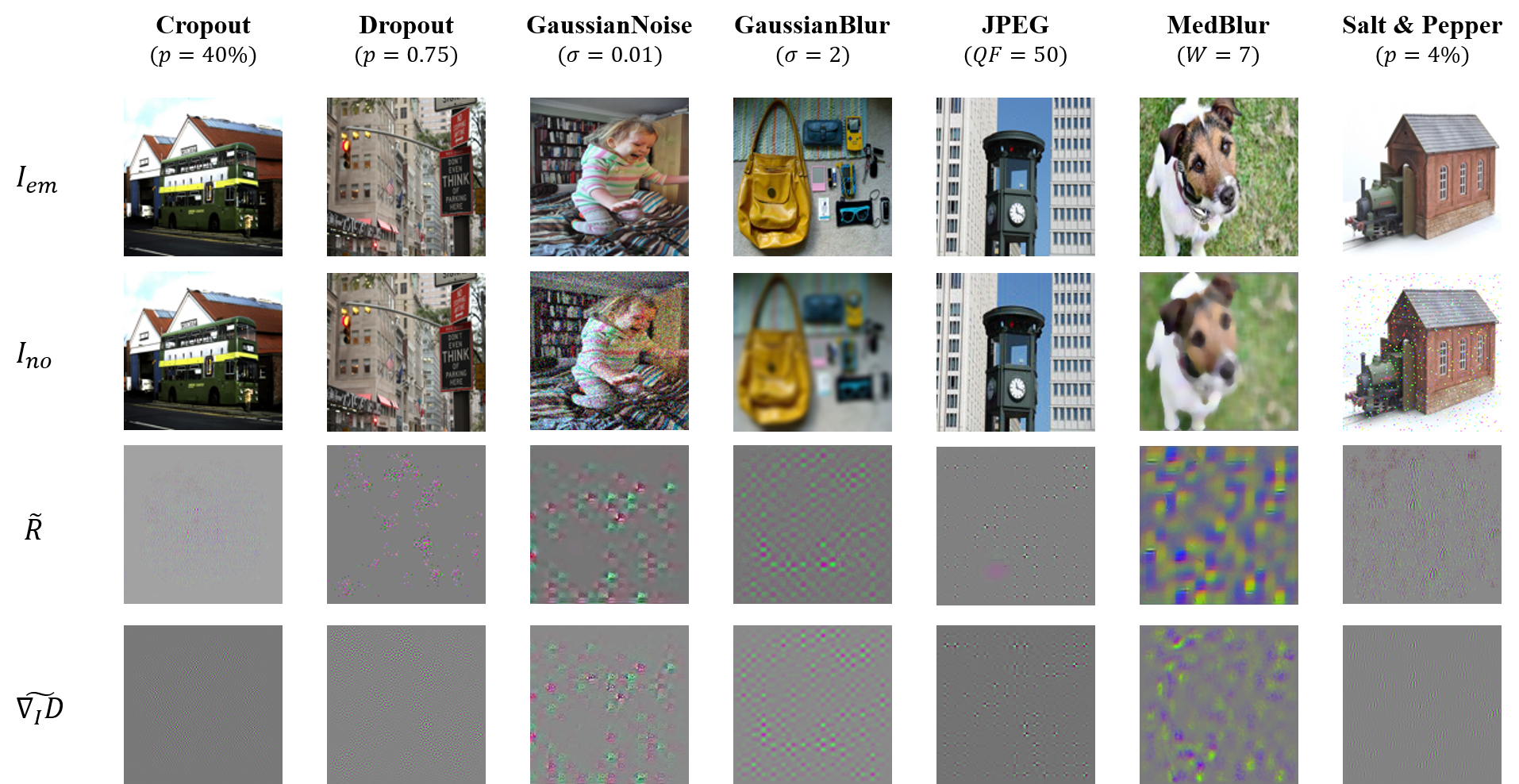}
        }
    \end{minipage}%
\caption{The distortions used for testing, and the corresponding encoded features for each distortion based on the proposed architecture.}
\label{Distortion}
\end{figure*}

In this section, we will show and discuss the visual quality and robustness of different methods for 7 types of distortions: ``Cropout'', ``Dropout'', ``Gaussian Noise'', ``Salt\&Pepper Noise'', ``Gaussian Blur'', ``Median Blur'' and ``JPEG Compression''. We visually show each distortion and the corresponding encoded features of the proposed architecture in Fig. \ref{Distortion}. The first row is the embedded images, the second row indicates the distorted images. The third row and fourth row represents the encoded features and the features that decoder needed respectively. We can see that for all the distortions, the encoder and decoder are coupling well which results in high consistency of the feature map. Besides, the encoded residual adaptively changed with the distortion, which indicates the great learning ability of the proposed architecture.

\captionsetup{font={footnotesize}}
\subsubsection{Cropout Distortion}
Cropout refers to the distortion that crops a block of image with a certain ratio out and replaces the cropped region with original images. In training stage, we fixed the parameter of cropout ratio as 40\%, and we test the cropped ratio from 10\% to 50\%. The experimental results are shown in Table \ref{table_VQ_cropout} and Table \ref{table_cropout}.
\begin{table}[h]
\caption{The PSNR values of each methods for cropout distortion.} \centering
\label{table_VQ_cropout}
\scalebox{0.9}{
\begin{tabular}{c|cccc}
\toprule[2pt]
{Methods} & {HiDDen \cite{zhu2018hidden}} & {TSDL \cite{liu2019a}} & {MBRS \cite{jia2021mbrs}} & {Proposed} \\ 
\midrule
{PSNR(dB)}  & {40.62} & {47.48} & {48.05}  & {\textbf{50.74}} \\
\bottomrule[2pt]
\end{tabular}
}
\end{table}

\begin{table}[h]
\caption{Extraction accuracy with different cropout ratios.} \centering
\label{table_cropout}
\scalebox{0.9}{
\begin{tabular}{c|ccccc}
\toprule[2pt]
{Ratio (\%)} & {10} & {20} & {30} & {40} & {50} \\ 
\midrule
{HiDDen \cite{zhu2018hidden}}  & {95.63\%} & {94.73\%} & {88.75\%} & {76.88\%} & {61.67\%}\\
\midrule
{TSDL \cite{liu2019a}}  & {98.72\%} & {98.54\%} & {96.88\%} & {93.75\%}& {93.21\%}\\
\midrule
{MBRS \cite{jia2021mbrs}}  & {99.71\%} & {99.22\%} & {97.18\%} & {90.43\%}& {83.50\%}\\
\midrule
{Proposed}  & {\textbf{100\%}} & {\textbf{100\%}} & {\textbf{99.51\%}} & {\textbf{97.28\%}}& {\textbf{91.21\%}}\\
\bottomrule[2pt]
\end{tabular}
}
\end{table}
As seen in Table \ref{table_VQ_cropout}, the proposed method will lead to the watermarked images with the highest PSNR values, which are at least 2dB larger than the compared schemes. Under such PSNR level, the proposed scheme still maintains the strongest robustness in different cropout ratios. For the crop ratio of 10\% to 40\%, the extraction accuracy of the proposed scheme is all higher than 97\%. And as the 
cropout ratio increasing, the advantage of the proposed algorithm is more and more obvious. 

\subsubsection{Dropout Distortion}
Dropout distortion indicates the operation that randomly replaces a certain ratio of image pixels with the original image. But different from cropout, such image pixels are randomly sampled in the whole image. We fixed the ratio of 40\%. And for testing, we change the ratio from 20\% to 60\%. The experimental results are shown in Table \ref{table_VQ_dropout} and Table \ref{table_dropout}.
\begin{table}[h]
\caption{The PSNR values of each methods for dropout distortion.} \centering
\label{table_VQ_dropout}
\scalebox{0.9}{
\begin{tabular}{c|cccc}
\toprule[2pt]
{Methods} & {HiDDen \cite{zhu2018hidden}} & {TSDL \cite{liu2019a}} & {MBRS \cite{jia2021mbrs}} & {Proposed} \\ 
\midrule
{PSNR(dB)}  & {42.59} & {53.59} & {58.63}  & {\textbf{58.97}} \\
\bottomrule[2pt]
\end{tabular}
}
\end{table}

\begin{table}[h]
\caption{Extraction accuracy with different dropout ratios.} \centering
\label{table_dropout}
\scalebox{0.9}{
\begin{tabular}{c|ccccc}
\toprule[2pt]
{Ratio (\%)} & {60} & {50} & {40} & {30} & {20} \\ 
\midrule
{HiDDen \cite{zhu2018hidden}}  & {82.71\%} & {86.74\%} & {87.08\%} & {89.58\%} & {90.21\%}\\
\midrule
{TSDL \cite{liu2019a}}  & {90.42\%} & {92.29\%} & {93.54\%} & {95.21\%}& {97.54\%}\\
\midrule
{MBRS \cite{jia2021mbrs}}  & {90.63\%} & {92.58\%} & {94.15\%} & {94.73\%}& {96.29\%}\\
\midrule
{Proposed}  & {\textbf{94.63\%}} & {\textbf{99.51\%}} & {\textbf{100\%}} & {\textbf{100\%}}& {\textbf{100\%}}\\
\bottomrule[2pt]
\end{tabular}
}
\end{table}
We can see in Table \ref{table_VQ_dropout}, the proposed framework maintains the best visual quality. Besides, Table \ref{table_dropout} indicates the superior performance of the proposed scheme compared with other frameworks. For all the dropout ratios, the proposed framework guarantees the best extraction accuracy, which is at least 2\% higher than the other schemes.

\subsubsection{Gaussian Noise}
Gaussian noise is commonly appeared in message transmission. For training stage, we randomly distort the image with a variance from 0.001 to 0.04. And the testing variance ranges from 0.01 to 0.05. The results are shown in Table \ref{table_VQ_GN} and Table \ref{table_GN}.
\begin{table}[h]
\caption{The PSNR values of each methods for Gaussian noise distortion.} \centering
\label{table_VQ_GN}
\scalebox{0.9}{
\begin{tabular}{c|cccc}
\toprule[2pt]
{Methods} & {HiDDen \cite{zhu2018hidden}} & {TSDL \cite{liu2019a}} & {MBRS \cite{jia2021mbrs}} & {Proposed} \\ 
\midrule
{PSNR(dB)}  & {36.25} & {39.46} & {39.70}  & {\textbf{40.13}} \\
\bottomrule[2pt]
\end{tabular}
}
\end{table}

\begin{table}[h]
\caption{Extraction accuracy with different variance of Gaussian noise.} \centering
\label{table_GN}
\scalebox{0.9}{
\begin{tabular}{c|ccccc}
\toprule[2pt]
{Variance} & {0.01} & {0.02} & {0.03} & {0.04} & {0.05} \\ 
\midrule
{HiDDen \cite{zhu2018hidden}}  & {89.58\%} & {86.46\%} & {83.96\%} & {83.12\%} & {79.17\%}\\
\midrule
{TSDL \cite{liu2019a}}  & {92.08\%} & {91.25\%} & {88.33\%} & {87.08\%}& {82.92\%}\\
\midrule
{MBRS \cite{jia2021mbrs}}  & {99.91\%} & {99.42\%} & {98.10\%} & {96.09\%}& {94.15\%}\\
\midrule
{Proposed}  & {\textbf{100\%}} & {\textbf{99.71\%}} & {\textbf{98.34\%}} & {\textbf{96.58\%}}& {\textbf{95.90\%}}\\
\bottomrule[2pt]
\end{tabular}
}
\end{table}
As seen in Table \ref{table_VQ_GN}, the proposed scheme maintains the highest PSNR value which is more than 40dB. From Table \ref{table_GN} we can see that, in all the testing variance, the proposed scheme maintains the best robustness against Gaussian noise distortion. The accuracy is significantly higher than HiDDen and TSDL, but for MBRS, the advantage appears in the large variance.

\subsubsection{Salt\&Pepper Noise}
Similar to Gaussian noise, Salt\&Pepper Noise is commonly appeared in transmission too, which randomly sampled a certain ratio of image pixels into a noise. The training ratio is randomly selected from 0.001 to 0.04. And the testing ratio ranges from 0.01 to 0.05. The final results are shown in Table \ref{table_VQ_SP} and Table \ref{table_SP}.
\begin{table}[h]
\caption{The PSNR values of each methods for Salt\&Pepper noise distortion.} \centering
\label{table_VQ_SP}
\scalebox{0.9}{
\begin{tabular}{c|cccc}
\toprule[2pt]
{Methods} & {HiDDen \cite{zhu2018hidden}} & {TSDL \cite{liu2019a}} & {MBRS \cite{jia2021mbrs}} & {Proposed} \\ 
\midrule
{PSNR(dB)}  & {46.04} & {51.16} & {51.79}  & {\textbf{52.43}} \\
\bottomrule[2pt]
\end{tabular}
}
\end{table}

\begin{table}[h]
\caption{Extraction accuracy with different variance of Salt\&Pepper noise.} \centering
\label{table_SP}
\scalebox{0.9}{
\begin{tabular}{c|ccccc}
\toprule[2pt]
{Ratio (\%)} & {0.01} & {0.02} & {0.03} & {0.04} & {0.05} \\ 
\midrule
{HiDDen \cite{zhu2018hidden}}  & {95.12\%} & {93.79\%} & {93.45\%} & {92.92\%} & {90.42\%}\\
\midrule
{TSDL \cite{liu2019a}}  & {97.29\%} & {95.63\%} & {93.54\%} & {92.71\%}& {91.46\%}\\
\midrule
{MBRS \cite{jia2021mbrs}}  & {98.05\%} & {98.74\%} & {98.34\%} & {97.56\%}& {96.68\%}\\
\midrule
{Proposed}  & {\textbf{99.41\%}} & {\textbf{99.51\%}} & {\textbf{99.22\%}} & {\textbf{99.12\%}}& {\textbf{98.73\%}}\\
\bottomrule[2pt]
\end{tabular}
}
\end{table}
We can see that the PSNR of the proposed scheme is higher than other compared schemes. For robustness, the proposed framework still ensures the best performance. The extraction accuracy of the proposed scheme is bigger than 98\% in all the test distortions, which indicates the strong robustness against Salt\&Pepper noise.

\subsubsection{Gaussian Blur}
For Gaussian blur distortion, we fixed the noise layer with variance of 2, and conduct the Gaussian blurring operation with variance from 0 to 2 in testing to show the robustness. The visual quality comparison and extraction accuracy are shown in Table \ref{table_VQ_GB} and Table \ref{table_GB}.
\begin{table}[h]
\caption{The PSNR values of each methods for Gaussian blur distortion.} \centering
\label{table_VQ_GB}
\scalebox{0.9}{
\begin{tabular}{c|cccc}
\toprule[2pt]
{Methods} & {HiDDen \cite{zhu2018hidden}} & {TSDL \cite{liu2019a}} & {MBRS \cite{jia2021mbrs}} & {Proposed} \\ 
\midrule
{PSNR(dB)}  & {46.21} & {45.97} & {47.91}  & {\textbf{48.41}} \\
\bottomrule[2pt]
\end{tabular}
}
\end{table}

\begin{table}[h]
\caption{Extraction accuracy with different variance of Gaussian blur.} \centering
\label{table_GB}
\scalebox{0.9}{
\begin{tabular}{c|cccc}
\toprule[2pt]
{Variance} & {0.0001} & {0.5} & {1} & {2} \\ 
\midrule
{HiDDen \cite{zhu2018hidden}}  & {95.44\%} & {95.21\%} & {94.33\%} & {84.37\%} \\
\midrule
{TSDL \cite{liu2019a}}  & {99.92\%} & {99.79\%} & {98.48\%} & {93.21\%}\\
\midrule
{MBRS \cite{jia2021mbrs}}  & {98.64\%} & {98.25\%} & {97.66\%} & {87.80\%}\\
\midrule
{Proposed}  & {\textbf{100\%}} & {\textbf{100\%}} & {\textbf{99.51\%}} & {\textbf{94.34\%}}\\
\bottomrule[2pt]
\end{tabular}
}
\end{table}
As seen in Table \ref{table_VQ_GB} and Table \ref{table_GB}, the proposed scheme produces the watermarked images with the highest PSNR value compared with other schemes. And the extraction accuracy is higher than the compared schemes too. 

\subsubsection{Median Blur}
Median Blur is a commonly used image processing operation. To train the robustness, we fixed the blurring window of $7\times7$ as the training parameter. And we test the robustness with blurring window of $3\times3, 5\times5$ and $7\times7$. The experimental results are shown in Table \ref{table_VQ_MB} and Table \ref{table_MB}.
\begin{table}[h]
\caption{The PSNR values of each methods for median blur distortion.} \centering
\label{table_VQ_MB}
\scalebox{0.9}{
\begin{tabular}{c|cccc}
\toprule[2pt]
{Methods} & {HiDDen \cite{zhu2018hidden}} & {TSDL \cite{liu2019a}} & {MBRS \cite{jia2021mbrs}} & {Proposed} \\ 
\midrule
{PSNR(dB)}  & {37.07} & {38.64} & {40.98}  & {\textbf{41.18}} \\
\bottomrule[2pt]
\end{tabular}
}
\end{table}
\vspace{-0.3cm}
\begin{table}[h]
\caption{Extraction accuracy with different windows of median blur.} \centering
\label{table_MB}
\scalebox{1}{
\begin{tabular}{c|ccc}
\toprule[2pt]
{Windows} & {$3\times3$} & {$5\times5$} & {$7\times7$} \\ 
\midrule
{HiDDen \cite{zhu2018hidden}}  & {86.25\%} & {83.70\%} & {79.71\%} \\
\midrule
{TSDL \cite{liu2019a}}  & {99.38\%} & {97.21\%} & {95.12\%} \\
\midrule
{MBRS \cite{jia2021mbrs}}  & {99.42\%} & {98.93\%} & {97.27\%} \\
\midrule
{Proposed}  & {\textbf{99.81\%}} & {\textbf{99.61\%}} & {\textbf{98.34\%}}\\
\bottomrule[2pt]
\end{tabular}
}
\end{table}
As seen in Table \ref{table_VQ_MB} and Table \ref{table_MB}, we ensure the best visual quality compared with other schemes. Besides, we maintain the highest extraction accuracy under the PSNR in Table \ref{table_VQ_MB}. For all the testing median blur windows, the proposed scheme achieves more than 98\% accuracy, which indicates superior robustness against median blur distortion.

\subsubsection{JPEG Compression}
JPEG compression always appears in image saving and format conversion. In training stage, we choose the noise layer with QF=50 which is proposed by MBRS \cite{jia2021mbrs} as the default parameter. And we test the JPEG compression attack with QF from 40 to 90 to show the robustness. The experimental results are shown in Table \ref{table_VQ_JPEG} and Table \ref{table_JPEG}. 
\begin{table}[h]
\caption{The PSNR values of each methods for JPEG compression distortion.} \centering
\label{table_VQ_JPEG}
\scalebox{0.9}{
\begin{tabular}{c|cccc}
\toprule[2pt]
{Methods} & {HiDDen \cite{zhu2018hidden}} & {TSDL \cite{liu2019a}} & {MBRS \cite{jia2021mbrs}} & {Proposed} \\ 
\midrule
{PSNR(dB)}  & {33.29} & {39.39} & {45.16}  & {\textbf{46.84}} \\
\bottomrule[2pt]
\end{tabular}
}
\end{table}
\vspace{-0.3cm}
\begin{table}[h]
\caption{Extraction accuracy with different quality factors of JPEG compression.} \centering
\label{table_JPEG}
\scalebox{0.9}{
\begin{tabular}{c|cccccc}
\toprule[2pt]
{QF} & {40} & {50} & {60} & {70} & {80} & {90} \\ 
\midrule
{HiDDen \cite{zhu2018hidden}}  & {86.67\%} & {91.24\%} & {92.92\%} & {93.33\%} & {93.54\%} & {94.38\%}\\
\midrule
{TSDL \cite{liu2019a}}  & {91.04\%} & {91.46\%} & {93.96\%} & {94.21\%} & {94.35\%} & {94.74\%}\\
\midrule
{MBRS \cite{jia2021mbrs}}  & {94.83\%} & {94.93\%} & {96.68\%} & {97.66\%} & {97.66\%} & {98.84\%}\\
\midrule
{Proposed}  & {\textbf{98.15\%}} & {\textbf{99.02\%}} & {\textbf{100\%}} & {\textbf{100\%}} & {\textbf{100\%}} & {\textbf{100\%}}\\
\bottomrule[2pt]
\end{tabular}
}
\end{table}
As seen in Table \ref{table_VQ_JPEG}, the proposed scheme ensures high quality watermarked images with PSNR value that is more than 46dB. For robustness, the proposed framework is significantly higher than the compared schemes, as shown in Table \ref{table_JPEG}. Especially for QF=50, the proposed scheme reaches more than 99\% accuracy, which implies that the algorithm can be effectively used in practice since in most cases, the quality factors will not be less than 50 in practical use. 

\subsection{Analysis of \textbf{De-END} Architecture}
In this section, we will conduct more analysis of the proposed \textbf{De-END} architecture to verify the design.
\subsubsection{Architecture Improvements}\label{AI}
\begin{figure}
    \begin{minipage}[p]{1\linewidth}
        \centering{
            \includegraphics[width=1\linewidth]{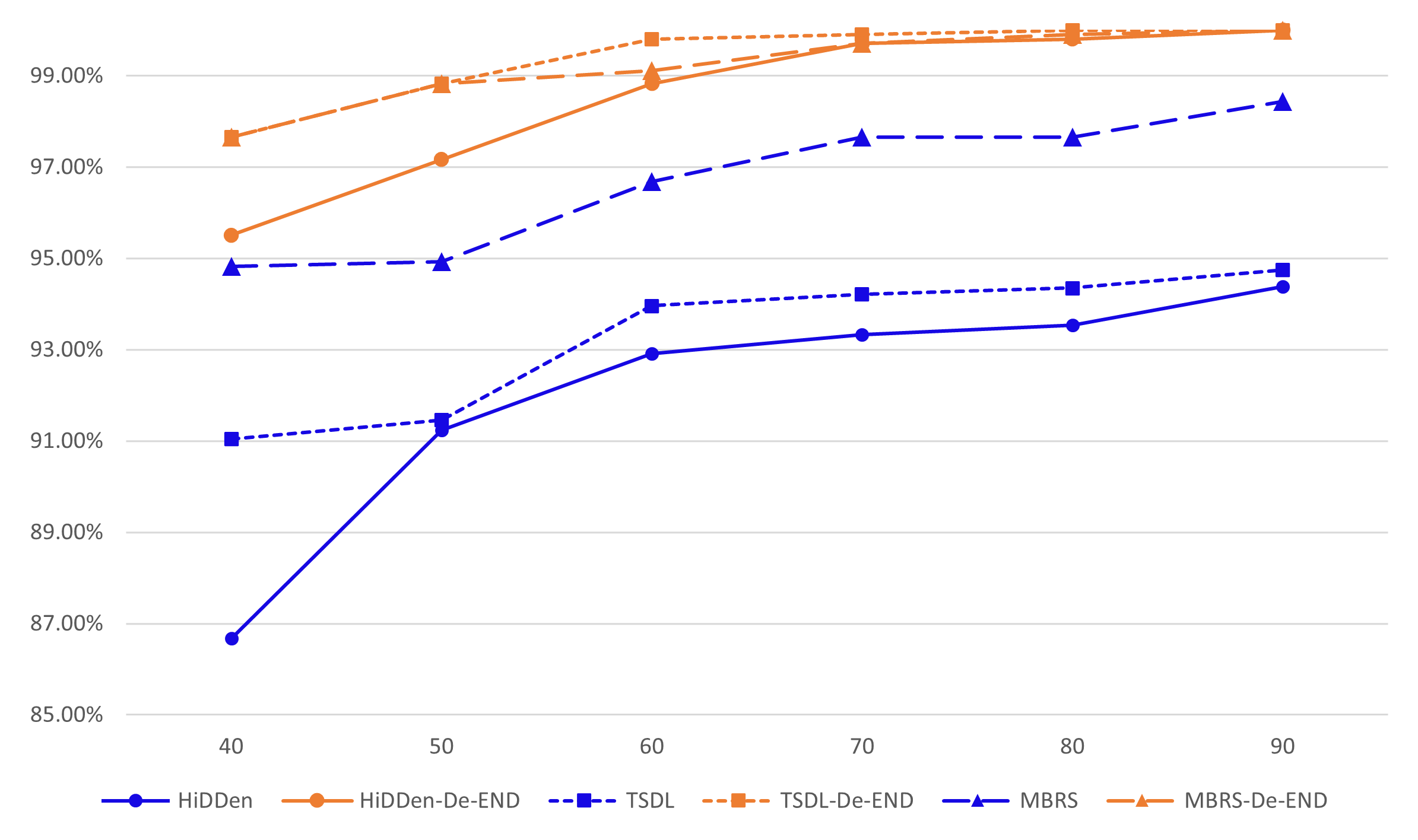}
        }
    \end{minipage}%
\caption{The robustness comparison with different architecture.}
\label{AS_archi}
\end{figure}

The key contribution of this paper is designing the \textbf{De-END} architecture, which achieves better robustness and visual quality compared with other schemes. The most important thing we should confirm is whether the performance improvement comes from the architecture or the encoder/decoder backbone. In order to verify the effectiveness of \textbf{De-END} architecture, we conduct the following experiments. We choose the same decoder as well as the noise layer proposed by HiDDen, TSDL and MBRS, and set them as the decoder in \textbf{De-END} architecture. As for encoder, we use the proposed encoder structure. Then we train the network with JPEG compression distortion. The experimental results are shown in Table \ref{table_AB_archi} and Fig. \ref{AS_archi}.
\begin{table}[h]
\caption{The PSNR values of each decoder backbones for JPEG compression distortion.} \centering
\label{table_AB_archi}
\scalebox{0.9}{
\begin{tabular}{c|ccc}
\toprule[2pt]
{Methods} & {HiDDen \cite{zhu2018hidden}} & {TSDL \cite{liu2019a}} & {MBRS \cite{jia2021mbrs}} \\ 
\midrule
{Original PSNR(dB)}  & {33.29} & {39.39} & {45.16}  \\
\midrule
{\textbf{De-END} PSNR(dB)}  & {45.62} & {45.68} & {45.57}  \\
\bottomrule[2pt]
\end{tabular}
}
\end{table}

\captionsetup{font={footnotesize}}
As shown in Table \ref{table_AB_archi}, the proposed De-END architecture has significantly improved the visual quality of methods HiDDen \cite{zhu2018hidden} and TSDL \cite{liu2019a}. For MBRS \cite{jia2021mbrs}, it achieves a little improvement. But for robustness, it can be seen that applying \textbf{De-END} architecture will greatly improve the extraction accuracy. As seen in  Fig. \ref{AS_archi}, the blue line indicates the extraction accuracy with the original \textbf{END} architecture, the orange line represents the extraction accuracy with \textbf{De-END} architecture. For all the methods, simply utilizing \textbf{De-END} architecture without any backbone changing will bring performance improvement. This indicates that the improvements mainly come from the architecture instead of the backbone design.

\subsubsection{Architecture Comparison}
As aforementioned, in \textbf{De-END}, two decoders share the same parameters. And we believe sharing parameter is the key to realizing feature coupling of encoder and decoder. In this section, we will test whether it is necessary to share the same parameter. We change the \textbf{De-END} architecture from ``Decoder-Encoder-NoiseLayer-Decoder'' (denoted as \textbf{De-END}) to ``Decoder$_A$-Encoder-NoiseLayer-Decoder$_B$'' (denoted as \textbf{De$_A$END$_B$}). Decoder$_A$ has the same structure as Decoder$_B$, but we do not force them to share parameters. If \textbf{De$_A$END$_B$}can achieve similar performance as \textbf{De-END}, it means that sharing the same parameters will not be helpful in feature coupling, 
only applying a ``Decoder-Encoder'' like structure in embedding stage is enough to improve the performance. In addition to ``Decoder$_A$-Encoder-NoiseLayer-Decoder$_B$'', we also test the structure of \textbf{END} where the encoder only adopts watermark as input, denote as ``\textbf{E$_W$ND}''. If ``\textbf{E$_W$ND}'' performs better, it means setting decoder before encoder is useless, the main improvements just rely on the input changing of the encoder. So to verify our design, we train \textbf{De-END}, \textbf{De$_A$END$_B$} and \textbf{E$_W$ND} respectively. After training, the PSNR values of \textbf{De$_A$END$_B$}, \textbf{E$_W$ND} and \textbf{De-END} are 46.40dB, 46.88dB and 46.84dB respectively. The extraction accuracy is shown in Table \ref{table_DD}.

\begin{table}[h]
\caption{Extraction accuracy with different architecture against JPEG compression.} \centering
\label{table_DD}
\scalebox{0.9}{
\begin{tabular}{c|cccccc}
\toprule[2pt]
{QF} & {40} & {50} & {60} & {70} & {80} & {90} \\ 
\midrule
{\textbf{De$_A$END$_B$}}  & {83.98\%} & {89.84\%} & {94.92\%} & {97.27\%} & {97.95\%} & {98.44\%}\\
\midrule
{\textbf{E$_W$ND}}  & {92.29\%} & {95.61\%} & {96.88\%} & {99.02\%} & {100\%} & {100\%}\\
\midrule
{\textbf{De-END}}  & {\textbf{98.15\%}} & {\textbf{99.02\%}} & {\textbf{100\%}} & {\textbf{100\%}} & {\textbf{100\%}} & {\textbf{100\%}}\\
\bottomrule[2pt]
\end{tabular}
}
\end{table}

It can be seen that under the same level of PSNR, \textbf{De-END} maintains the best extraction accuracy compared with \textbf{De$_A$END$_B$} and \textbf{E$_W$ND}, which means sharing the parameters with two decoders is a good way to couple encoder and decoder. Besides, setting the decoder before encoder is also necessary for getting better performance. It is worth noting that maybe training \textbf{De$_A$END$_B$} in some way may realize a similar performance as \textbf{De-END}, but we believe sharing parameters in these two decoders is a fast and effective way to realize good performance.

\subsubsection{Backbone of Decoder}
As illustrated in Section \ref{AI}, different structures of decoder will result in similar performance. Since the most commonly used union in HiDDen \cite{zhu2018hidden}, TSDL \cite{liu2019a} and MBRS \cite{jia2021mbrs} is ``Conv-BN-ReLU'', in this section, we will explore the relationship between the number of ``Conv-BN-ReLU'' union and the performance. We train the structure of the decoder with 3 to 7 ``Conv-BN-ReLU'' union and test the performance. The corresponding results are shown in Table \ref{table_AB_VQ_decoder} and Table \ref{table_AB_decoder}.
%the architecture of \textbf{De-END} will surely improve the performance. And we can also see in Fig. \ref{AS_archi} that the
\begin{table}[h]
\caption{The PSNR values of each decoder backbones for JPEG compression distortion.} \centering
\label{table_AB_VQ_decoder}
\scalebox{1}{
\begin{tabular}{c|ccccc}
\toprule[2pt]
{Number} & {3} & {4} & {5} & {6} & {7}\\ 
\midrule
{PSNR(dB)}  & {45.35} & {45.94} & {45.58} & {45.05} & {45.62}  \\
\bottomrule[2pt]
\end{tabular}
}
\end{table}

\begin{table}[h]
\caption{Extraction accuracy with different decoder backbones against JPEG compression.} \centering
\label{table_AB_decoder}
\scalebox{0.9}{
\begin{tabular}{c|cccccc}
\toprule[2pt]
{QF} & {40} & {50} & {60} & {70} & {80} & {90} \\ 
\midrule
{3 union}  & {78.91\%} & {83.11\%} & {88.28\%} & {93.46\%} & {97.41\%} & {99.81\%}\\
\midrule
{4 union}  & {91.60\%} & {93.95\%} & {97.17\%} & {98.63\%} & {99.81\%} & {100\%}\\
\midrule
{5 union}  & {94.43\%} & {97.46\%} & {98.44\%} & {99.71\%} & {100\%} & {100\%}\\
\midrule
{6 union}  & {95.80\%} & {96.97\%} & {98.05\%} & {99.41\%} & {100\%} & {100\%}\\
\midrule
{7 union}  & {95.51\%} & {97.17\%} & {98.83\%} & {99.71\%} & {100\%} & {100\%}\\
\bottomrule[2pt]
\end{tabular}
}
\end{table}
It can be seen from Table \ref{table_AB_VQ_decoder} that the PSNR will not significantly change as the number of ``Conv-BN-ReLU'' union increases, and will maintain stability at a high level of 45dB. But from Table \ref{table_AB_decoder} we can conclude that the extraction accuracy grows as the number of ``Conv-BN-ReLU'' union increases from 3 to 5. But for 5 to 7 unions, the performance becomes similar. It indicates that deepening the network will not necessarily produce better performance. Applying 5 to 7 ``Conv-BN-ReLU'' union is enough to ensure great performance.

\subsubsection{Backbone of Encoder}
In the proposed De-END architecture, the encoder adopts a very simple structure that contains one three ``Up-Conv-BN-ReLU''  blocks. Such structure has already shown its powerful performance, and deepening the network may not result in better performance. So in this section, we mainly investigate the influence of the up-sampling ways. We choose three commonly used upsampling ways: ``Un-pooling'', ``Transpose-conv'' and ``Nearest-interpolating'' to test the performance. Specifically, we only change the up-sampling ways in ``Up-Conv-BN-ReLU'' union and further train each network respectively. For ``Un-pooling'', since it is necessary to give the max-pooling location for each feature map in order to realize ``Un-pooling'' operation, we randomly sample the location and fixed it in training and testing. The final PSNR with ``Un-pooling'', ``Transpose-conv'' and ``Nearest-interpolating'' backbone are 46.53dB, {46.48dB} and 46.84dB respectively. The extraction accuracy are shown in  Table \ref{table_AB_encoder}.
%  visual quality and Table \ref{table_AB_VQ_encoder} and
% \begin{table}[h]
% \caption{The PSNR values of each decoder backbones for JPEG compression distortion.} \centering
% \label{table_AB_VQ_encoder}
% \scalebox{0.8}{
% \begin{tabular}{c|ccc}
% \toprule[2pt]
% {Up-sampling Rule} & {Un-pooling} & {T-conv} & {N-interpolating} \\ 
% \midrule
% {PSNR(dB)}  & {46.53} & {46.48} & {46.84} \\
% \bottomrule[2pt]
% \end{tabular}
% }
% \end{table}

\begin{table}[h]
\caption{Extraction accuracy with different up-sampling rules against JPEG compression.} \centering
\label{table_AB_encoder}
\scalebox{0.9}{
\begin{tabular}{c|cccccc}
\toprule[2pt]
{QF} & {40} & {50} & {60} & {70} & {80} & {90} \\ 
\midrule
{Un-pooling}  & {89.36\%} & {94.04\%} & {95.31\%} & {97.85\%} & {98.93\%} & {99.81\%}\\
\midrule
{T-conv}  & {97.27\%} & {98.83\%} & {99.41\%} & {100\%} & {100\%} & {100\%}\\
\midrule
{N-interpolating}  & {98.15\%} & {99.02\%} & {100\%} & {100\%} & {100\%} & {100\%}\\
\bottomrule[2pt]
\end{tabular}
}
\end{table}
% As seen in Table \ref{table_AB_VQ_encoder}, 
Different up-sampling rules will result in similar PSNR values of the watermarked images. As for robustness, ``Transpose-conv'' and ``Nearest-interpolating'' maintain similar high-level extraction accuracy, but for ``Un-pooling'', the performance will be worse than the other two rules. We conclude the reason that when we conduct ``Un-pooling'' operation, we fixed the max-pooling location for each feature map, and the randomly sampled location may not be optimal, which will result in bad performance. In general, applying ``Transpose-conv'' and ``Nearest-interpolating'' all can ensure great performance.

\section{Conclusion}
The existing DNN-based watermarking algorithms mainly adopt an \textbf{END} backbone which contains an encoder, a noise layer and a decoder. In this paper, we designed a novel decoder-driven DNN-based watermarking network dubbed \textbf{De-END}. The motivation comes from the potential drawbacks of the existing \textbf{END} framework we discovered, that is, encoder may embed redundant features which is not necessary for decoding into the image. And we deeply investigate the reason as the encoder and decoder cannot be well coupled under \textbf{END} framework. To the best of our knowledge, we are the first to give the analysis of such drawback and we hope it will benefit the follow-up work. As for mechanism, in order to address such problem, we propose the architecture of \textbf{De-END}, which cascades the decoder before the encoder. Based on this framework, the features of encoder and decoder can be better shared, so encoder and decoder can be better coupled. Various experiments show that the visual quality and watermark robustness of the proposed architecture is significantly better than the existing state-of-the-art algorithms, which greatly proves the superior performance of the \textbf{De-END} architecture.
{
\bibliographystyle{IEEEtran}
\bibliography{TMM}
}

\end{document}

\title{Supplementary Images}

\captionsetup{font={footnotesize}}
\begin{figure*}
    \begin{minipage}[t]{1\linewidth}
        \centering{
            \includegraphics[width=0.95\linewidth]{LaTeX/ExperimentResults_D.png}
        }
    \end{minipage}%
\caption{The robustness tests on traditional image processing distortions. We show the visual quality of original image $I_o$, the enhanced image $I_{en}$ and the distorted image $I_{no}$ attacked by eight different types of traditional distortions: Cropout, Dropout, Gaussian Noise, Gaussian Blur, JPEG Compression, Medium Blur, Resize and Salt$\and$Pepper.}
\label{ExperimentalResults_D}
\end{figure*}

\captionsetup{font={footnotesize}}
\begin{figure*}
    \begin{minipage}[t]{1\linewidth}
        \centering{
            \includegraphics[width=0.95\linewidth]{LaTeX/ExperimentResults_ND.png}
        }
    \end{minipage}%
\caption{The robustness tests on non-differentiable distortions. We show the visual quality of original image $I_o$, the enhanced image $I_{en}$ and the distorted image $I_{no}$ influenced by five different types of non-differentiable distortions: style transfer, instant message app transmission and screen-shooting.}
\label{ExperimentalResults_ND}
\end{figure*}